%% file: main_v7.tex
\documentclass[letterpaper,11pt]{article}
\usepackage[utf8]{inputenc}

\usepackage[style=chem-rsc,articletitle=true]{biblatex}
\usepackage[margin=1in]{geometry}
\usepackage{amsmath, amsfonts, amssymb, amsthm}
\usepackage{bm}
\usepackage{graphicx}
\usepackage[colorlinks=true,allcolors=blue]{hyperref}
\usepackage{authblk}
\graphicspath{{./}{figures/}{supp/}}

\begin{document}
\title{Improving Jet A-1 Thermal-Oxidative Stability through Selective Removal of Unwanted Trace Species via 3.7 \AA{} Chabazite Filtration}

\author[1]{Morteza Roostaeinia}
\author[2,3]{Ehsan Alborzi}
\author[4]{Xue Yong}
\author[1,5]{Kyungwha Park}
\author[1,5,6]{Vsevolod Ivanov}
\affil[1]{Department of Physics, Virginia Tech, Blacksburg, VA 24061, USA}
\affil[2]{Energy Innovation Centre, The University of Sheffield, Sheffield S9 1ZA, UK}
\affil[3]{School of Mechanical, Aerospace and Civil Engineering, The University of Sheffield, Sheffield S1 4DT, UK}
\affil[4]{Department of Electrical Engineering and Electronics, The University of Liverpool, Liverpool L69 3GJ, UK}
\affil[5]{Virginia Tech Center for Quantum Information Science and Engineering, Blacksburg, VA 24061, USA}
\affil[6]{Virginia Tech National Security Institute, Blacksburg, VA 24060, USA}
\date{}
\maketitle

\include{v7_abstract.tex}
\include{v7_introduction.tex}

\include{v7_methods.tex}

\include{v7_discussion.tex}

\section{Acknowledgements}
This work was supported by the Horizon 2020--Clean Sky 2 programme under research grant agreement 150089. The authors acknowledge the University of Liverpool High Performance Computing (HPC) services and access to ARCHER2 via MCC. X.Y. gratefully acknowledges funding from the Leverhulme Trust. We acknowledge Advanced Research Computing (ARC) at Virginia Tech for providing computational resources and technical support that contributed to the results reported in this paper (\url{https://arc.vt.edu/}). The authors also acknowledge support from Virginia Tech startup funds. We further acknowledge financial support from the Leverhulme Trust.

\printbibliography

\end{document}

%% file: v7_abstract.tex
\begin{abstract}
The thermal stability of Jet A-1 fuel is strongly affected by trace heteroatomic species that promote thermal oxidative deposit formation, as well as antioxidant additives such as butylated hydroxytoluene, which preserve fuel stability. 3.7~\AA{} chabazite is a tunable microporous adsorbent, but optimizing its composition requires balancing promoter removal against antioxidant loss. Here, we use density functional theory and  \textit{ab initio} molecular dynamics (AIMD) to evaluate this trade-off using two compositional descriptors: framework acidity (Si/Al $= 35$--$8$) and bimetallic substitution (Co, Mg, Zn) at fixed Al content. Lowering Si/Al strengthens adsorption of all species, including BHT, confirming an intrinsic selectivity penalty for acidity-only tuning. In contrast, bimetallic substitution introduces chemically selective behavior, strengthening uptake of specific promoters while either suppressing or enhancing antioxidant adsorption depending on dopant identity. AIMD simulations at 400~K further show that adsorption energy alone is insufficient to describe calculated trends, because molecular mobility inside chabazite can influence residence time and effective capture during fuel treatment. Analysis of the density of states and the participating wavefunctions reveals signatures that are consistent with dopant-dependent adsorption shifts. These results establish a unified adsorption--transport--electronic screening framework for selecting chabazite compositions that remove deposit promoters while preserving antioxidant functionality in Jet A-1 treatment.
\end{abstract}

%% file: v7_introduction.tex
\section{Introduction}
Jet A-1 fuel is vulnerable to thermal oxidative degradation when bulk fuel temperatures are in the range of 100 to 300 $^{\circ}$C, leading to insoluble products and surface deposits that can foul fuel-system components and compromise reliability \cite{Hazlett1991,Jia2021,Alborzi2024}. This degradation is strongly influenced by trace heteroatomic constituents and contaminants, including reactive organosulfur species \cite{An2026}, aromatic nitrogen compounds, oxygenated species, and metal-organic complexes \cite{Zabarnick2019}. These contaminants participate in radical oxidation pathways, form surface-active intermediates, and accelerate deposit growth even when present at low concentrations \cite{Hazlett1991, Taylor1976, TaylorFrankenfeld1978, Parks2022}. Because modern jet fuels, in particular sustainable aviation fuels, may vary significantly in composition, antioxidant content, and contaminant load, treatment strategies must address both chemical reactivity and the selective removal of the most deleterious minor species.

Adsorptive fuel treatment has emerged as a practical route to improve thermal-oxidative stability by reducing concentrations of deposit promoters upstream from hot-section components \cite{Alborzi2024,Velu2003}. In previous work, treatment of Jet A-1 fuel using monoliths coated with 3.7 \AA{} chabazite (CHA) reduced the rate of oxidative surface deposition; the effective reduction depended on framework acidity (Si/Al), monolith size, and whether the filtration was combined with hybrid activated carbon treatment \cite{Alborzi2024, Alborzi2019, Alborzi2020, Alborzi2022}. 

Those results established CHA as an effective platform for selective filtering of unwanted trace polar species, but also revealed a design trade-off: among the CHA-only treatments, the Si/Al $= 25$ monolith produced the largest reduction in surface-deposition propensity, whereas the greatest improvement in thermal-oxidative stability was obtained using a Si/Al $= 17$ CHA/activated-carbon combined bed \cite{Alborzi2024}. A further unresolved observation from the previous study was that, dibutyl disulfide (DBDS) showed limited removal of less than $5\%$ by the Si/Al $= 17$ CHA monolith in the doped model-fuel experiment, despite its relatively strong static DFT adsorption. The earlier work tentatively attributed this discrepancy to the narrow, linear geometry of DBDS, which could permit rapid passage through the monolith without sufficient contact time for effective capture, and identified molecular-dynamics analysis as necessary for further investigation \cite{Alborzi2024}. 

In this work, we use first-principles simulations to optimize CHA performance by tuning framework acidity and local electronic structure through variation of the Si/Al ratio and substitution of metal heteroatoms. The Si/Al ratio controls the K$^{+}$ Lewis acid site density and framework charge density, while heteroatom substitution perturbs local electronic structure \cite{Palcic2020,PerezBotella2022,Bordiga2005}. A transferable design framework is needed to couple these chemistry controls with molecule--pore size/shape compatibility and intrapore transport, so that promoter removal can be improved without excessive antioxidant loss \cite{Yang2018}.

\begin{figure}[t]
    \centering
    \includegraphics[width=\columnwidth,trim={6.0cm 1.2cm 6.0cm 0.2cm},clip]{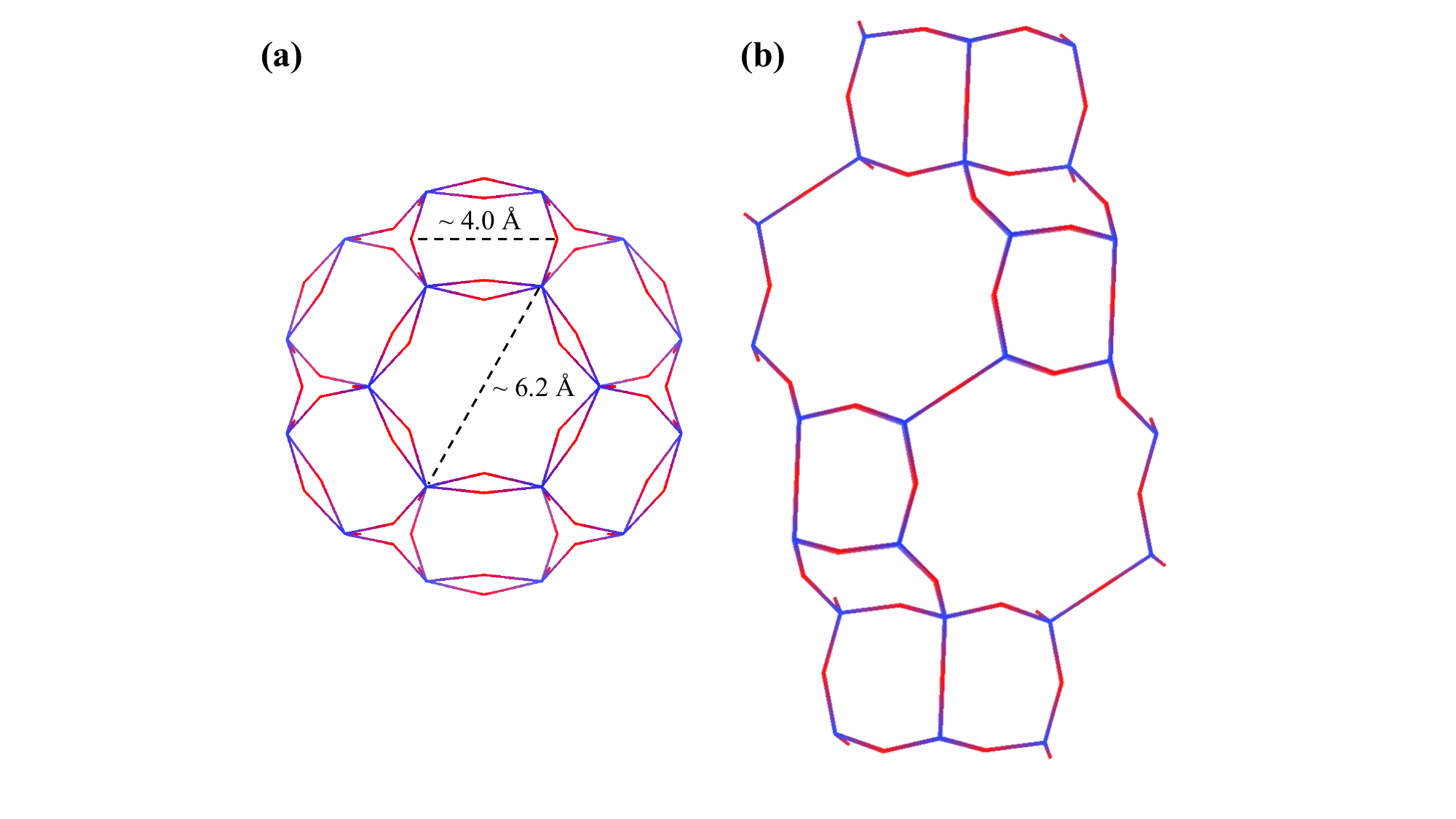}
    \caption{Framework topology of chabazite (CHA). (a) Top view of the cage--window motif. The dashed lines indicate an O--O distance of approximately 4.0~\AA{} across an 8MR window and a Si--Si distance of approximately 6.2~\AA{} across a \textit{cha} cage. (b) Side view of the connected CHA framework.}
    \label{fig:CHA_topology}
\end{figure}

Using the CHA model shown in Fig.~\ref{fig:CHA_topology}, we evaluate compositions spanning Si/Al $= 35$--$8$ and bimetallic variants containing Co, Mg, and Zn. The limited number of bimetallic elements was chosen due to the availability of doped CHA structures in the database \cite{IZA-CHA}.  For each composition, we calculate adsorption energies for five representative Jet A-1-relevant species: aniline, DBDS, Fe-naphthenate, ethanol, and butylated hydroxytoluene (BHT). Additionally, for each species we calculate apparent self-diffusion coefficients at 400~K in the most acidic framework using ab initio molecular dynamics (AIMD) simulations to assess how molecular size and interactions with the CHA framework influence mobility inside the pores \cite{allen2017_comp_sim_liq}. We find that decreasing the Si/Al ratio strengthens adsorption for all species, including BHT, whereas dopant identity enables more selective behavior. Analysis of the projected density of states further links these adsorption shifts to dopant-dependent electronic effects. Together, these results provide design guidance for improving promoter removal while minimizing antioxidant depletion.

%% file: v7_methods.tex
\section{Methods}

\subsection{Electronic structure calculations}
The quantum-chemistry simulations were performed using the CP2K/Quickstep package \cite{cp2k, quickstep} with periodic boundary conditions applied in all three spatial directions. Long-range electrostatics were handled within the Generalized Poisson Solver formalism. Geometry optimizations used the Broyden--Fletcher--Goldfarb--Shanno (BFGS) algorithm \cite{optimization_bfgs} within the density functional theory framework.

Electronic exchange--correlation effects were described using the Perdew--Burke--Ernzerhof (PBE) functional \cite{gga} in combination with Goedecker--Teter--Hutter pseudopotentials \cite{Goedecker1996} and DZVP-MOLOPT-SR-GTH basis sets \cite{VandeVondele2007_MOLOPT} within the mixed Gaussian/plane-wave formalism \cite{Lippert1997}. A plane-wave cutoff of 350~Ry and a relative cutoff of 50~Ry were used for the charge-density representation. Dispersion interactions were included using Grimme's D3 correction \cite{Grimme2010_D3} with Becke--Johnson damping \cite{Grimme2011_D3BJ}. Spin-polarized DFT was used only for systems containing Co and/or Fe atoms. Self-consistent field convergence employed the orbital transformation method with a conjugate-gradient minimizer. the inner and outer SCF convergence thresholds were both set to $5 \times 10^{-6}$.

The chabazite (CHA) framework is a small-pore zeolite built from TO$_4$ tetrahedra (T-sites) that share corners, where $T$=Si or Al. It contains large \textit{cha} cages and double six-ring (\textit{d6r}) units. The \textit{cha} cages are connected through windows formed by eight-membered rings (8MRs) with an effective aperture of approximately 3.7~\AA{} \cite{PerezBotella2022,IZA_ZeoliteDatabase}. These windows limit molecular access to the cages and provide the size-selective behavior relevant to Jet A-1 fuel treatment. The framework topology (IZA code: CHA), consisting of a 36 T-site rhombohedral unit cell ($\mathrm{Si_{36}O_{72}}$ in its pure-silica form), was obtained from the IZA Structure Commission database \cite{IZA_ZeoliteDatabase}. As illustrated in Fig.~\ref{fig:CHA_topology}, panel (a) shows the cage--window motif and the characteristic dimensions of the 8MR opening and the \textit{cha} cage, whereas panel (b) shows a side view of the connected CHA framework.
 
To probe the effect of varying framework Al content, we used the 36 T-site CHA unit cell and generated compositions $\mathrm{K}_x\mathrm{Si}_{36-x}\mathrm{Al}_x\mathrm{O}_{72}$ with $x = 1$--4, corresponding to Si/Al ratios of 35, 17, 11, and 8. Charge neutrality was maintained by adding $x$ extra-framework $\mathrm{K}^+$ cations. For each Al content, all symmetry-distinct Al arrangements were generated and fully optimized; the most stable configuration at each $x$ was used for subsequent adsorption calculations. This set of chabazite structures is henceforth referred to as ``mono-Al.''

Increasing framework Al content (lower Si/Al ratio) can enhance adsorption of polar promoters but may also counterproductively increase the adsorption of antioxidants such as hindered phenols, which could reduce the thermal oxidative stability of the treated fuel. This drawback can be overcome by introducing framework co-dopants (e.g., divalent metals), which can create new adsorption environments that alter selectivity among chemically distinct species \cite{Daouli2023}. Bimetallic frameworks (CHA-Al$_2$Mg, CHA-Al$_2$Co, CHA-Al$_2$Zn) were generated by replacing one Si atom in the CHA-Al$_2$ configuration with a divalent metal dopant ($\mathrm{Mg}^{2+}$, $\mathrm{Co}^{2+}$, or $\mathrm{Zn}^{2+}$), resulting in frameworks containing two $\mathrm{Al}^{3+}$ and one $\mathrm{M}^{2+}$ heteroatom with 4 $\mathrm{K}^{+}$ atoms. To accommodate adsorbate molecules, the structure was duplicated along the $c$-axis. All five species were modeled in vacuum. Adsorption energies were computed as
\begin{equation}
E_{\rm ad} = E_{\rm total} - E_{\rm framework} - E_{\rm adsorbate},
\label{eq:Ead}
\end{equation}
with negative values indicating stronger bonding \cite{Piccini2015}. For each adsorbate--framework combination, multiple initial configurations were sampled by varying the molecular orientation and adsorption site, followed by geometry optimization. Depending on the adsorbate size and framework composition, 3--13 initial configurations were tested. The lowest-energy configuration from each set was then used to determine adsorption energy. 

For the lowest-energy configurations, local adsorption geometries were characterized using minimum-image distances between adsorbate functional groups and nearby framework atoms/cations. The descriptors include heteroatom--cation distances, hydrogen-bond distances and angles, ring-centroid--cation distances for aromatic adsorbates, and metal/carboxylate contacts for Fe-naphthenate.

\subsection{Ab initio molecular dynamics simulations}
To assess relative mobility and transport limitations for the five selected adsorbates inside CHA pores, ab initio molecular dynamics simulations were performed using CP2K/Quickstep at 400 K representing typical thermal oxidative conditions in Jet A-1 fuel systems \cite{Alborzi2024, Kuprowicz2004, SmitMaesen2008}. Simulations were carried out for the Si/Al = 8 CHA framework. This composition was chosen because it provides the strongest adsorption environment within the mono-Al series. The optimized adsorption geometry with the lowest energy for each molecule was used as the initial structure.

The AIMD simulations were performed in the NVT ensemble using a canonical sampling velocity rescaling (CSVR) thermostat with a time constant of 750 fs. The same PBE functional with D3(BJ) dispersion, GTH pseudopotentials, and DZVP-MOLOPT-SR-GTH basis sets used for the static DFT calculations were employed for the MD force evaluations. A timestep of 1.0 fs was used, and each trajectory was propagated for 90 ps at 400 K. 

Self-diffusion coefficients $D_s$ were extracted from the mean squared displacement using the Einstein relation \cite{allen2017_comp_sim_liq, Frenkel2002_mol_sim}:
\begin{equation}
    D_s = \lim_{t\rightarrow\infty} \frac{1}{6t} \left\langle \vert \bm{r}(t)-\bm{r}(0) \vert^2 \right\rangle,
\end{equation}
where $\left\langle \vert \bm{r}(t)-\bm{r}(0) \vert^2 \right\rangle$ denotes the time dependent mean squared displacement of the adsorbate. Linear fits were performed over the linear region of the mean-squared displacement (MSD) curves; all fits had regression values $R^2 \geq 0.94$. The resulting apparent self-diffusion coefficients $D_\mathrm{s}$ are used to compare relative molecular mobility among the adsorbates inside CHA. Given the finite trajectory lengths accessible to \textit{ab initio} MD, these values should be interpreted as order-of-magnitude indicators of relative intrapore transport rather than converged equilibrium diffusion coefficients.

\subsection{Projected density of states and HOMO--LUMO wavefunction analysis}

Projected density of states (PDOS) calculations were performed for optimized bare CHA frameworks, including the mono-Al reference structures and the bimetallic CHA variants. The PDOS was used to identify framework- and dopant-derived states near the band edges and to assess how Al content and divalent dopants modify the electronic structure of the adsorption environment. Cube files for the highest occupied molecular orbital (HOMO) and lowest unoccupied molecular orbital (LUMO) were also generated for selected optimized bare frameworks to visualize their spatial distributions. These analyses characterize the electronic structure of the bare adsorption sites rather than direct adsorbate--framework charge transfer. Spin-polarized DFT was used for Co-containing frameworks.

%% file: v7_discussion.tex
\begin{figure*}[t]
    \centering
    \includegraphics[width=\textwidth]{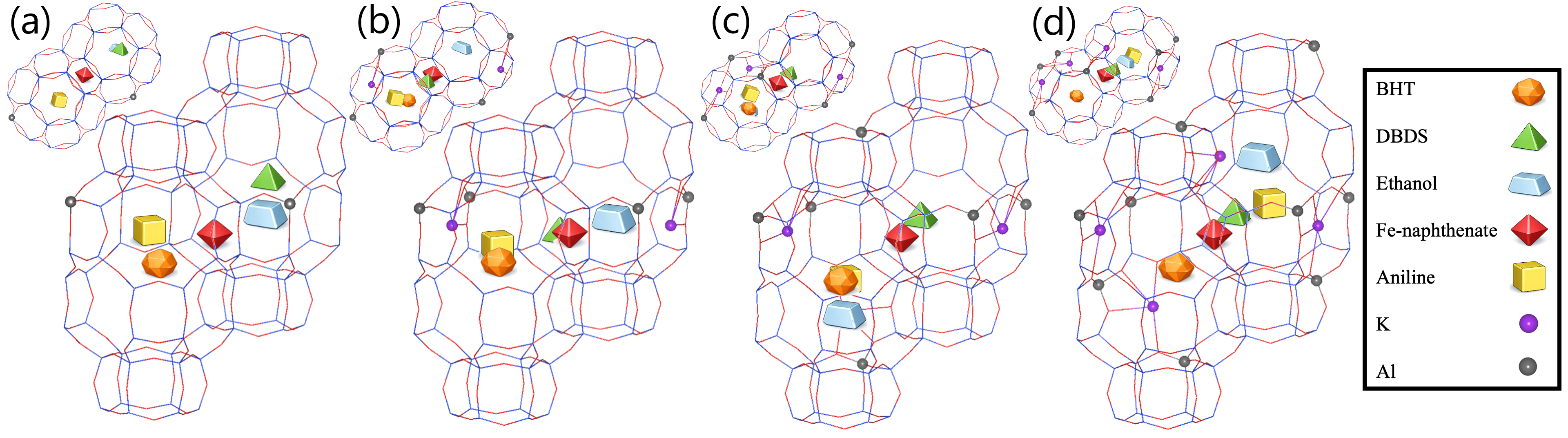}
    \caption{Representative lowest-energy adsorption locations of BHT, DBDS, ethanol, Fe-naphthenate, and aniline in mono-Al CHA frameworks with increasing Al content: (a) Si/Al = 35, (b) Si/Al =17, (c) Si/Al = 11, and (d) Si/Al = 8. Each main panel shows a side view of the doubled CHA framework, and the inset shows the corresponding top view. Colored shapes mark the positions of the independently optimized minimum-energy configurations for each adsorbate. Gray and purple spheres denote framework Al sites and extra-framework K$^+$ cations, respectively.}
    \label{fig:adsorption_locations}
\end{figure*}

\section{Results and Discussion}

\subsection{Effect of Aluminum Content on Adsorption Energies \label{sec31}} 
Before discussing the adsorption results, it is important to clarify the nature of the acidity descriptor used in this work. In H-form zeolites, the Si/Al ratio controls Br{\o}nsted acidity through the density of bridging Si--OH--Al sites. In the K-exchanged CHA frameworks studied here, however, no Br{\o}nsted acid sites are present: each Al$^{3+}$ substitution introduces one unit of negative framework charge that is compensated by a K$^{+}$ extra-framework cation rather than a proton. The Si/Al ratio therefore serves as a quantitative descriptor for the K$^{+}$ Lewis acid site density: with Si/Al = 35, 17, 11, and 8 corresponding to 1, 2, 3, and 4 K$^{+}$ cations per 36 T-site unit cell, respectively. All five adsorbates form their closest contacts with the K$^{+}$/framework-O environment, with no short direct contacts to Al or substituted metal sites in the mono-Al series (See Table~S11 in the Supporting Information \cite{supplement}). Therefore, the K$^{+}$ site density, equivalently expressed through the Si/Al ratio, provides a quantitative descriptor for the Lewis acid character of the adsorption environment relevant to polar species capture.

We first investigate the effect of acidity on adsorption selectivity by systematically varying the Si/Al ratio of CHA (35, 17, 11, and 8). Representative lowest-energy adsorption locations for the five adsorbates in the Al doped CHA series are shown in Fig.~\ref{fig:adsorption_locations}. Each adsorbate marker represents an independently optimized minimum-energy configuration containing a single species molecule. These overlays provide a qualitative map of preferred adsorption regions within the CHA cages and near charge-compensating sites. This reveals several qualitative trends across the mono-Al CHA series. The large species, DBDS and Fe-naphthenate, have preferred binding sites in between the two large $cha$ cages that do not have a strong dependence on the Si/Al ratio, likely due to the size of these molecules. The smaller aniline and BHT molecules instead prefer to adsorb near the center of a $cha$ cage, while the lowest-energy binding site of the highly polar ethanol adsorbate shifts away from the center of the large $cha$ cage with increasing Al content.

\begin{figure*}[t]
    \centering
    \includegraphics[width=\textwidth]{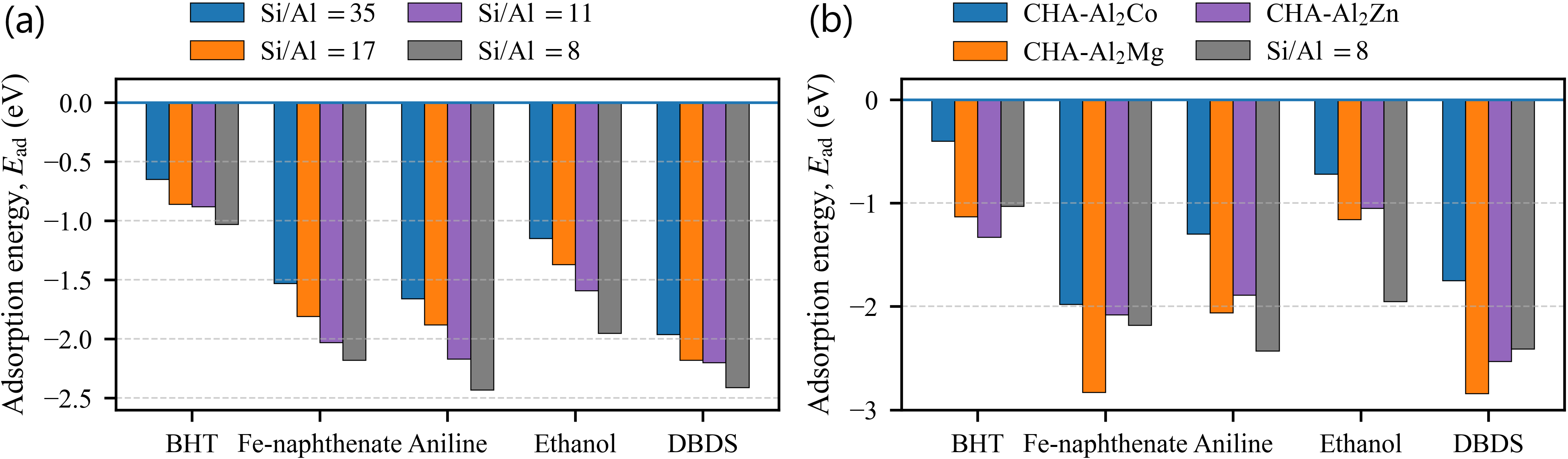}
    \caption{(a) Adsorption energies of representative species on mono-Al CHA as a function of framework Si/Al ratio. (b) Adsorption energies on bimetallic CHA--Al$_2$M variants (M= Co, Mg, Zn), compared with the Si/Al = 8 reference. }
    \label{fig:adsorption_overview}
\end{figure*}

Fig.~\ref{fig:adsorption_overview}(a) shows the adsorption energies as the framework Al content increases (i.e., as the Si/Al ratio decreases). Decreasing the Si/Al ratio from 35 to 8 strengthens adsorption for all species, consistent with enhanced framework polarity and a higher density of Al--K$^{+}$ adsorption sites. The lowest-energy configurations show stronger binding at the highest Al loading for BHT, Fe-naphthenate, DBDS, aniline, and ethanol, although the magnitude of the change is molecule-dependent. This also reveals a selectivity trade-off. Higher Al content can improve uptake of deposit-promoting species, which is desirable for removing them through fuel adsorption treatment, but it can also increase adsorption of the antioxidant BHT. Prior experimental work considered the adsorption properties of chabazite frameworks with Si/Al ratios near $\sim 17$; our calculations find that between Si/Al = 17 and 11, BHT and DBDS adsorption remains nearly constant, whereas the binding energies of the polar Fe-naphthenate, ethanol, and aniline molecules continue to increase. Detailed adsorption energies are provided in Tables~S1--S10 of the Supporting Information \cite{supplement}. Percent changes discussed in the main text were calculated from the lowest-energy configurations \cite{supplement}.
In the mono-Al series, local contact analysis confirms that all five adsorbates interact primarily with the K$^{+}$/framework-O environment. Aniline preferentially orients its amino group toward framework O/K$^{+}$ sites, with N-H$\cdots$O contacts of 2.07--2.74 \AA{} and N$\cdots$K$^{+}$ distances of 2.89--3.28 \AA{}. Ethanol binds through its hydroxyl oxygen, forming O$\cdots$K$^{+}$ contacts of 2.68--2.84 \AA{}. BHT interacts via its phenolic-OH/aromatic region, with the closest O-H$\cdots$O contacts of 2.42--2.63 \AA{}. DBDS adsorbs with its sulfur-containing region and adjacent C-H groups oriented toward the pore wall, with S$\cdots$O contacts of 2.95--3.33 \AA{} and S$\cdots$K$^{+}$ distances of 3.22--6.81 \AA{}. Fe-naphthenate binds through its Fe/carboxylate region, with distances of 2.47--3.02 \AA{} between Fe and framework oxygens, and distances of 2.72--2.80 \AA{} between framework and molecule oxygens. A full summary of local contact descriptors across the mono-Al and bimetallic series is provided in Table S11 \cite{supplement}. These observations confirm that lowering the Si/Al ratio strengthens adsorption by increasing the density of Al-associated K$^{+}$/framework-O binding environments, while the specific local adsorption motif remains strongly dependent on the molecular functional groups.

\subsection{Impact of Bimetallic Doping on Adsorption Performance \label{sec32}}
Bimetallic substitution introduces an additional tuning parameter beyond simply increasing Al content. In these systems, one framework Si atom in the CHA-Al$_2$ structure is replaced by a divalent dopant (Mg$^{2+}$, Co$^{2+}$, or Zn$^{2+}$), giving CHA--Al$_2$M frameworks charge-compensated by four K$^+$ cations. Mg, Co, and Zn were chosen as representative divalent dopants to examine how changing the metal affects adsorption under comparable charge conditions. Although these structures do not have the same Si/Al ratio as the mono-Al $\mathrm{Si/Al}=8$ framework, they have the same total framework charge and the same number of charge-compensating K$^+$ cations. Therefore, Fig.~\ref{fig:adsorption_overview}(b) compares the bimetallic adsorption energies with the $\mathrm{Si/Al}=8$ mono-Al framework as a charge-equivalent high-charge-density reference.

CHA--Al$_2$Mg selectively strengthens only one of the deposit promoters considered here. DBDS binding increases by $\approx 18\%$, whereas aniline, ethanol, and Fe-naphthenate adsorption weaken relative to the Si/Al = 8 reference by $\approx 15\%$, $\approx 41\%$, and $\approx 12\%$, respectively. This selective response is consistent with Mg$^{2+}$ modifying the local electrostatic environment and the arrangement of nearby K$^+$/framework-O adsorption site.

CHA--Al$_2$Zn leaves the binding of Fe-naphthenate and DBDS mostly unaffected, while significantly weakening the absorption of aniline by $\approx 0.5$ eV and ethanol by $\approx 0.9$ eV. This behavior is consistent with Zn$^{2+}$ favoring soft Lewis acid interactions with sulfur in DBDS and metal coordination in Fe-naphthenate \cite{Parks2022,Hessou2019}. 
In contrast, CHA--Al$_2$Co is more selective; it weakens adsorption of most species, with the exception of Fe-naphthenate, which remains essentially unchanged.

For the antioxidant BHT, the dopants produce clearly divergent effects. CHA--Al$_2$Mg only slightly increases the BHT binding energy (by $\approx 10\%$), while CHA--Al$_2$Zn significantly increases it  by $\approx 29\%$, which could risk depletion. Conversely, CHA--Al$_2$Co substantially suppresses BHT adsorption ($\approx$61\% decrease), which may help reduce unintended BHT uptake. These trends suggest that bimetallic substitution can help tune promoter adsorption and BHT uptake more selectively than acidity only tuning.

\begin{figure*}[t]
    \centering
    \includegraphics[width=\textwidth]{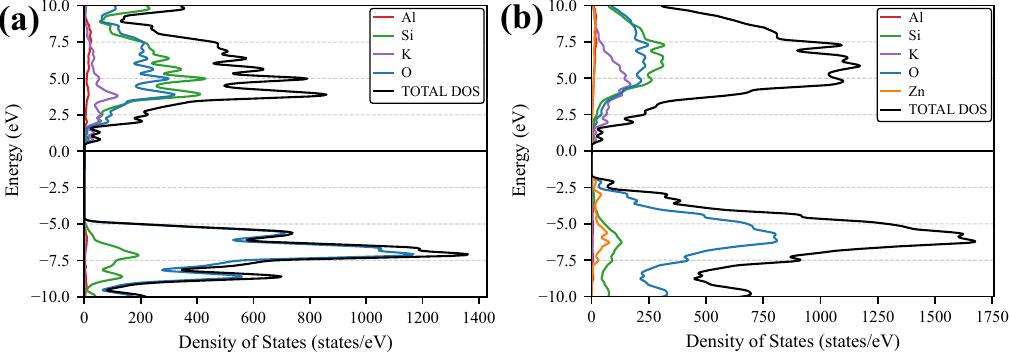}
    \caption{(a) Projected density of states (PDOS) for CHA--Al$_2$ (reference). (b) PDOS for CHA--Al$_2$Zn. Energies are referenced to the Fermi level ($E_F = 0$ eV).}
    \label{fig:dos_overview}
\end{figure*}

The projected densities of states (PDOS) shown in Fig.~\ref{fig:dos_overview}(a,b) provide insight into the origin of these effects. The reference PDOS for CHA--Al$_2$ in Fig.~\ref{fig:dos_overview}(a) shows that the valence band is dominated by O-$2p$ states, with limited contributions from Al-derived states near the Fermi level. In CHA--Al$_2$Zn, the Zn-$3d$ states introduce additional features near the band edge, consistent with the modified adsorption behavior of the Zn-doped framework. HOMO--LUMO wavefunction analysis of the optimized bare frameworks further supports this interpretation (See Table~S12 in the Supporting Information \cite{supplement}). 

\begin{figure}[!htb]
\centering
\includegraphics[width=0.95\textwidth]{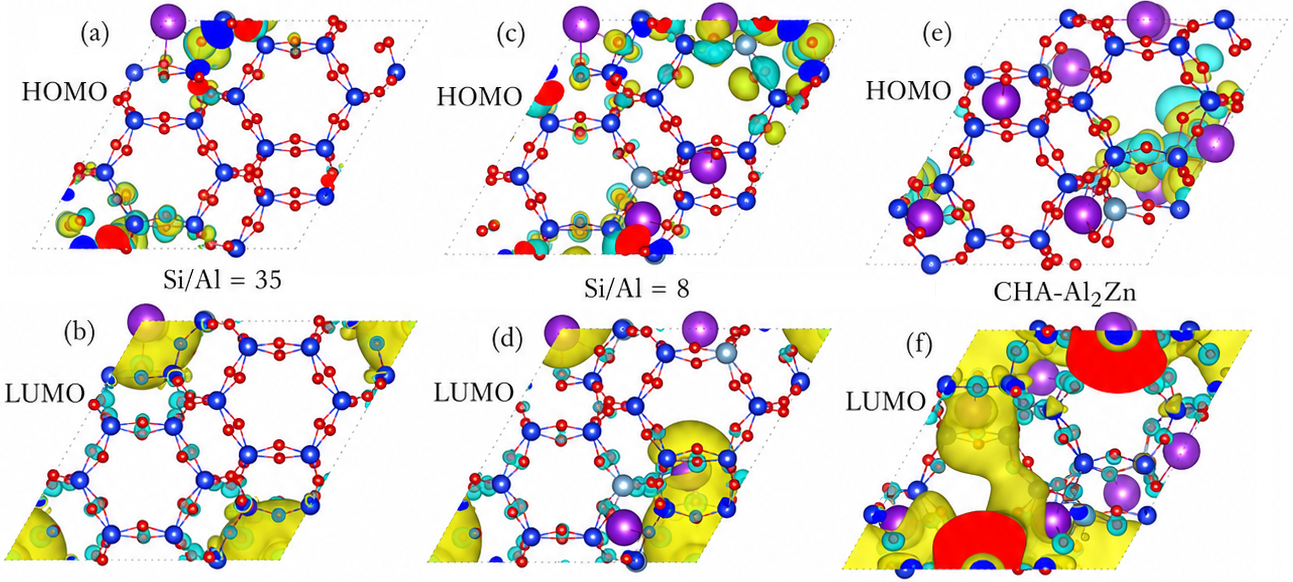}
\caption{
HOMO and LUMO wavefunction isosurfaces of representative optimized bare CHA frameworks.
Panels (a,b) show the low-Al mono-Al framework with Si/Al = 35, panels (c,d) show the high-Al mono-Al framework with Si/Al = 8, and panels (e,f) show the Zn-doped bimetallic framework CHA-Al$_2$Zn.
Yellow and cyan isosurfaces denote opposite phases of the orbital wavefunction.
The plots visualize the spatial distribution of band-edge states in the bare adsorption environments.
}
\label{fig:homo_lumo}
\end{figure}

Figure~\ref{fig:homo_lumo} shows the HOMO and LUMO wavefunction isosurfaces for representative optimized bare CHA frameworks.
The mono-Al frameworks retain relatively large HOMO--LUMO gaps, changing from 5.70 eV for Si/Al = 35 to 5.17 eV for Si/Al = 8. Therefore, the stronger adsorption at lower Si/Al is not primarily associated with a large global narrowing of the HOMO--LUMO gap, but rather with the increased density of polar Al/K$^+$ adsorption environments within the CHA cages. In contrast, Zn substitution reduces the HOMO--LUMO gap to 2.89 eV and produces more localized band-edge orbital density near the doped region, indicating that divalent metal substitution introduces electronically active framework sites. These electronic perturbations contribute to the species-specific selectivity observed in the adsorption energies.

Absolute energies are summarized in Table S1--S10 (Supporting Information) \cite{supplement}. Overall, bimetallic substitution shifts from acidity-driven global strengthening to chemically selective tuning, enabling tailored adsorption performance for Jet A-1 fuel treatment.

\subsection{Diffusion Kinetics and the Adsorption--Diffusion Trade-off \label{sec33}}
To complement the static DFT adsorption energies, we performed  MD simulations at 400~K for the most acidic CHA framework (Si/Al = 8). This temperature represents typical thermal oxidative conditions in Jet A-1 fuel systems. Diffusion coefficients $D_s$ were extracted from the mean-squared displacement using the Einstein relation (see Methods for details). Linear fits in the diffusive regime yielded $R^{2} \ge 0.94$ over the 90~ps trajectories \cite{KrishnaVanBaten2008}.

\begin{table}[ht]
    \centering
    \caption{ Apparent self-diffusion coefficients ($D_s$) of adsorbed molecules in the CHA framework at 400 K ($\mathrm{Si/Al}=8$), estimated from finite ab initio MD trajectories using the Einstein relation. Values represent relative molecular mobility within the simulation timescale and should be interpreted as order-of-magnitude transport estimates rather than converged equilibrium self-diffusion coefficients. The Fe-naphthenate trajectory is shorter than those of the other adsorbates owing to the additional computational cost of spin-polarized DFT.}
    \vspace{0.5cm}
    {\renewcommand{\arraystretch}{1.3}
    \begin{tabular}{ c | c }
        \hline
        \hline
        Molecule  & $D_s$ (cm²/s) \\
        \hline
        Fe-naphthenate & $4.84 \times 10^{-9}$ \\
        DBDS & $4.48 \times 10^{-8}$ \\
        aniline & $2.95 \times 10^{-8}$ \\
        BHT & $2.14 \times 10^{-8}$ \\
        ethanol & $1.91 \times 10^{-8}$ \\
        \hline
        \hline
    \end{tabular}
    }
    \label{diff_coef}
\end{table}

The apparent self-diffusion coefficients $D_s$ values (Table~\ref{diff_coef}) reveal substantial variation in relative molecular mobility across adsorbates. Fe-naphthenate exhibits the lowest mobility, consistent with its strong binding and bulky naphthenate structure that restricts passage through the 3.7~\AA{} 8MR windows. In contrast, DBDS displays the highest mobility, enabled by its linear and flexible chain that allows rapid transport through the CHA channels. Aniline, BHT, and ethanol show intermediate mobilities.

Notably, despite being the smallest molecule in the set, ethanol shows lower apparent mobility than aniline, consistent with its unusually short O$\cdots$K$^{+}$ contacts of 2.68--2.84~\AA{} (Table~S11)---the shortest adsorbate--cation contacts observed in this study---which localize ethanol near K$^{+}$ sites and reduce its effective cage-to-cage transport. We also note that the Fe-naphthenate trajectory is shorter than those of the other four adsorbates due to the additional computational cost of spin-polarized DFT; its extremely low MSD over the available simulation time is qualitatively consistent with strong binding and large molecular size, and the conclusion that Fe-naphthenate exhibits markedly lower mobility than all other species studied here is robust.
The MD results provide a plausible transport explanation for the apparent discrepancy between the static adsorption energies and prior experimental observations for monoliths coated with CHA \cite{Alborzi2024}. Although DBDS binds strongly in the DFT adsorption calculations, its larger apparent mobility suggests a shorter residence time near adsorption sites, which can reduce effective capture during fuel treatment under flow conditions. By contrast, the much lower apparent mobility of Fe-naphthenate suggests longer contact with the framework, which can enhance retention even when adsorption energy alone is not a complete descriptor. These trends indicate that adsorption strength alone is insufficient to predict filtration performance, and that molecular mobility should also be considered.

\subsection{Integrated Discussion \label{sec34}}
Selective fuel treatment in Jet A-1 is fundamentally a three-way optimization problem: maximize uptake of deposit promoters, minimize unintended antioxidant loss, and maintain transport properties that support retention under flow-through conditions. Our results show that the three-axis design space for CHA: acidity (Si/Al ratio), bimetallic dopant identity, and adsorbate size/shape relative to the 3.7~\AA{} window, can effectively moderate deposit promoters while retaining the antioxidant BHT.

Across the Al doped CHA series, decreasing the Si/Al ratio strengthens adsorption for all species, consistent with increased framework polarity and stronger host--species interactions. The magnitude of this strengthening is molecule-specific. For instance BHT adsorption varies significantly for different Si/Al ratios and bimetallic dopant compositions, highlighting an intrinsic selectivity challenge: although higher acidity improves capture of deposit promoters, it also binds the antioxidant more strongly which may still reduce the net thermal-oxidative stability of the treated fuel.

Bimetallic substitution provides an additional tuning lever beyond acidity. At fixed Al content (Si/Al = 8), divalent dopants (Mg, Zn, Co) modulate adsorption energetics in a molecule-specific manner. Doping the chabazite framework with Mg selectively strengthens DBDS and Fe-naphthenate binding while weakening adsorption of the other promoters, Zn slightly enhances BHT adsorption while decreasing binding of some promoters, wheras Co weakens all promoters except Fe-naphthenate and strongly suppresses BHT adsorption. These trends demonstrate that dopant identity can decouple promoter removal from antioxidant retention, shifting the material response from acidity-driven global strengthening to chemically selective tuning.

The five adsorbates naturally divide into groups that respond differently to these design axes, and identifying this grouping connects compositional choices to target molecule class. Ethanol and aniline show the steepest sensitivity to framework acidity due to their small size and polar nature (Fig.~\ref{fig:adsorption_overview}(a)), with binding energies that continue to increase from Si/Al = 11 to 8 even when larger molecules plateau. Their dimensions are comfortably within the 3.7~\AA{} 8MR aperture, so their capture is governed primarily by electrostatic interactions with the K$^{+}$/framework-O environment. This makes framework acidity the appropriate design parameter to tune the adsorption of this group. BHT occupies an intermediate position: it also binds through a polar phenolic-OH group, but its bulky di-tert-butyl substituents increase its effective size in the cage and moderate its mobility ($D_s = 2.14 \times 10^{-8}$ cm$^{2}$/s, Table~1), making it susceptible to non-selective adsorption and therefore the central selectivity challenge in the system. DBDS and Fe-naphthenate are more strongly influenced by pore-window compatibility and intrapore transport. As noted in Section \ref{sec31}, these two molecules prefer binding sites between the large $cha$ cages with limited Si/Al sensitivity, consistent with size-restricted access to the charge-compensating K$^{+}$ sites inside the cage interior. Their transport behavior diverges markedly despite both being large species. The flexible linear DBDS chain enables relatively high mobility ($D_s = 4.48 \times 10^{-8}$ cm$^{2}$/s), which may lead to rapid breakthrough under flow conditions despite its strong static binding energy. On the other hand, the rigid bulky structure of Fe-naphthenate severely restricts its mobility ($D_s = 4.84 \times 10^{-9}$ cm$^{2}$/s), favoring prolonged cage residence and effective retention even where its adsorption energy is not the highest in the series. These molecule-specific results highlight the importance of considering the polarity and size of target species when designing CHA composition for selective adsorption.

Adsorption energies alone are not enough to predict performance under realistic flow-through conditions. Although the apparent self-diffusion coefficients $D_s$ carry uncertainty from the finite ab initio MD trajectory lengths, the relative mobility ordering across the all five adsorbates is robust and consistent with the molecular size, shape, and binding characteristics discussed above. Fe-naphthenate exhibits much lower apparent mobility, suggesting longer residence near the framework, whereas DBDS shows higher apparent mobility despite strong static binding. This contrast highlights the role of molecular size/shape and pore-window compatibility in determining effective capture under flow. The combined adsorption--diffusion picture therefore refines material-selection criteria: optimal CHA frameworks must provide favorable thermodynamics for target promoters together with moderate diffusivity to avoid rapid breakthrough or pore saturation.

Electronic-structure signatures from PDOS and HOMO--LUMO wavefunction analyses supply a mechanistic basis for the observed dopant-dependent selectivity and close the interpretation loop between composition and performance. Across the mono-Al frameworks, the valence band remains dominated by O-$2p$ and framework-derived states, while increasing Al content mainly increases the density of polar Al/K$^+$ adsorption environments. This explains why adsorption strengthens with decreasing Si/Al even though the HOMO--LUMO gap changes only modestly across the mono-Al series. Divalent dopants produce a stronger electronic effect by introducing near-edge states and reducing the HOMO--LUMO gap, indicating the formation of electronically active doped regions. These dopant-induced electronic features provide an additional contribution to adsorption selectivity beyond the acidity-driven electrostatic effect.

Overall, these results establish a predictive materials-design strategy wherein framework acidity sets the baseline binding strength and targeted bimetallic substitution tunes chemical selectivity for promoters and antioxidants. This computational framework provides a molecular-level interpretation for the experimentally observed improvements in deposit control and antioxidant retention achieved with CHA-based monoliths in Jet A-1 fuel treatment \cite{Alborzi2024}.

\subsection{Summary and Design Implications \label{sec35}}
Periodic DFT and ab initio MD simulations were used to evaluate a central design challenge in Jet A-1 treatment with CHA sorbents: selective removal of deposit promoters without excessive loss of antioxidant functionality. Two composition controls were examined. First, decreasing Si/Al ratio strengthened adsorption for all species, confirming that acidity-driven tuning improves promoter capture but also increases BHT uptake. Second, bimetallic substitution at fixed Al content (Si/Al = 8) enabled selective tuning: Mg selectively strengthened Fe-naphthenate DBDS adsorption while weakening the other promoters, whereas Zn selectively strengthened BHT binding. Co strongly suppressed BHT adsorption while suppressing the binding of all promoters except Fe-naphthenate. MD simulations at 400 K further indicate that adsorption energy alone is insufficient to describe the calculated trends, because differences in molecular mobility inside CHA can affect residence time and effective capture during flow-through fuel treatment. PDOS and HOMO--LUMO wavefunction analyses further linked these trends to dopant-dependent electronic perturbations near the band edge, providing a mechanistic basis for the observed composition-controlled selectivity.

Taken together, these results establish a predictive, molecular-level framework for designing CHA-based sorbents that achieve selective removal of deposit promoters while preserving antioxidant functionality in Jet A-1 fuel treatment. The findings provide a computational foundation for the experimental improvements reported with CHA-coated monoliths and open the door to future studies involving competitive multi-component adsorption and fully adsorbate-projected electronic-structure analysis.